\documentclass[aps,prd,preprint,showpacs,amsmath,amssymb,amsfonts]{revtex4}

\def\vct#1{\mathbf{#1}}

\allowdisplaybreaks

\begin{document}

\title{Comment on recent papers regarding next-to-leading order
spin-spin effects in gravitational interaction}

\author{Jan Steinhoff and Gerhard Sch\"afer}
\affiliation{Theoretisch-Physikalisches Institut,
Friedrich-Schiller-Universit\"at,
Max-Wien-Platz~1, 07743 Jena, Germany}

\date{\today}

\begin{abstract}
It is argued that the tetrad in a recent paper by Porto and Rothstein
on gravitational spin-spin coupling should not have the given form.
The fixation of that tetrad was suggested by Steinhoff, Hergt, and
Sch\"afer as a possible source for the
disagreement found in the spin-squared dynamics. However, this
inconsistency will only show up in the next-to-leading order
spin-orbit dynamics and not in the spin-squared dynamics.
Instead, the disagreement found at the next-to-leading order
spin-squared level is due to a sign typo in the spin-squared paper by
Porto and Rothstein.
\end{abstract}

\vspace{2ex}
\noindent

\maketitle

\vspace{2ex}

In recent papers, Steinhoff, Hergt, and Sch\"afer derived the next-to-leading order (NLO) spin-squared dynamics
for binary black holes \cite{SHS08,HS08}. The result led to a spin-precession equation which disagreed with
an earlier result by Porto and Rothstein (PR) \cite{PR08b} based on the formalism of \cite{PR08a}.
The suggestion was given way in \cite{SHS08} that a different choice of the
tetrad $e_a^{\mu}$ may cure the disagreement. It is well known that a tetrad is only
fixed by the metric up to a local Lorentz transformation,
which made it a plausible source for the disagreement.
Both the correct fixation of the tetrad and the disagreement in
the spin-squared dynamics will be clarified here.

By evaluating, say for particle with index 1, $u_1^{\mu}u^{\nu}_1
g_{\mu\nu}$ in both the local and coordinate frames, one gets the
consistency condition (metric signature -2 as in the papers by PR)
\begin{equation}
( \tilde{v}^{a=0}_1 )^2 - \tilde{v}^{a=i}_1\tilde{v}^{a=i}_1  =
g_{00}(\vct{x}_1) + 2 g_{0i}(\vct{x}_1) v^i_1
+ g_{ij}(\vct{x}_1) v^i_1 v^j_1 \label{consist} \,,
\end{equation}
where $\tilde{v}^a_1=e^{a}_{\mu}(\vct{x}_1)u^{\mu}_1$ relates to the local frame
and $v^i_1=u^i_1$, $u^0_1=1$ to the coordinate frame.
This condition is not fulfilled for Eqs.\ (22) and (23) in \cite{PR08a}, using
leading order terms of the regularized metric for spinning binary black holes in harmonic coordinates
on the right-hand side of this condition. This inconsistency will show up at the NLO spin-orbit dynamics
via the spin supplementary condition $S^{0i}_1 \tilde{v}^{a=0}_1 + S^{ij}_1 \tilde{v}^{a=j}_1 = 0$.
In passing we note that a choice consistent with (\ref{consist}) and
sufficient for the NLO is given by
\begin{align}
\tilde{v}^{a=0}_1 &= 1 -  \frac{G_Nm_2}{r} \,,\\
\tilde{v}^{a=i}_1 &= v^i_1 + \frac{G_Nm_2}{r} (v^i_1 - 2 v^i_2) +  \frac{G_N}{r^2} S_2^{ij}n^j\,. \label{vi}
\end{align}
The $v^i_2$ term in (\ref{vi}) arises from the boosted Schwarzschild metric in harmonic coordinates.
Here the tetrad $e_a^{\mu}$ was obtained from Eq.\ (34) in \cite{P06}
(notice $e^I_{\mu}=e^a_{\mu}\Lambda^I_a$), which is the fixation
of the tetrad that enters the derivation of the Feynman rules.

The disagreement at the spin-squared level is indeed not due to a further
modification (via a local Lorentz transformation) of the tetrad.
Instead, by comparing Eq. (73) in \cite{PR08a} with Eq. (62) in \cite{PR08b},
a simple calculation reveals that the signs of the last terms are not the same.
By redoing the corresponding calculations one can check that Eq. (73) in \cite{PR08a} is correct,
i.e., there is a sign typo in (62) of \cite{PR08b}.
The spin-precession equations of \cite{PR08b} and \cite{SHS08} now coincide after the spin transformation
\begin{equation}
 \vct{S}^{\text{NW}}_1 = \vct{S}^{\text{SHS}}_1 
	- \frac{1}{2 m_1^3}  [ ( \vct{P}_1 \times \vct{S}_1 ) \times \dot{\vct{P}}_1 ] \times \vct{S}_1 \label{trafo} \,,
\end{equation}
has been performed, where the index ``SHS'' refers to the spin expression in
\cite{SHS08} and ``NW'' to the corresponding one in \cite{PR08b}.
$\vct{P}_1$ differs from $\vct{p}_1$ in \cite{SHS08} only by higher order terms,
cf., Eq.\ (5) in \cite{SHS08}. However in this multisheeted domain, it was
quite a cumbersome task to check the correctness of the other
terms (taking for granted the correctness of the Feynman-diagram expressions)
or to deduce the reason for the disagreement by comparing with our result.

It should be noted that Newton-Wigner (NW) variables are originally only
defined in flat spacetime, and that a generalization to curved
spacetime is not unique. In our understanding NW variables should have a
standard canonical meaning
\footnote{Our understanding of NW variables
emphasizes a slightly different aspect from their original setting \cite{NW49}, where
locality of the position variables, not canonicity of the independent variables,
is the defining property. However, in flat spacetime the canonicity of the variables
follows from the former and we think this is the important property that should be promoted to
curved spacetime.}, i.e.,
\begin{align}
\{ z^i_a, P_{a j} \} &= \delta_{ij} \,, \\
\{ S_{a(i)}, S_{a(j)} \} &= \epsilon_{ijk} S_{a(k)} \,,
\end{align}
{\em zero otherwise}, which is true in our papers. In the papers by PR the NW variables
are constructed such that, besides agreement with the usual NW variables in flat spacetime,
the spin has constant length, i.e., the spin equation of motion manifestly describes a spin precession.
While at the spin-orbit and spin(1)-spin(2) level this implies that
the spin is also standard canonical, this is not true at the spin(1)-spin(1)
level; see Eq.\ (13) in \cite{SHS08}.
Indeed, the spin-squared term in Eq.\ (\ref{trafo}) is not related to a
canonical transformation and should in our understanding, where NW
stands for ``standard canonical'', be included into the definition of the NW variables.
However, the spin equations of motion in \cite{PR08b} and \cite{SHS08}
are physically equivalent, so the discrepancy in the understanding of
NW variables is a matter of taste only.

A comparison of the center-of-mass motion is still missing. This is
necessary because all
$\vct{S}_1^2$ terms in the potential do not contribute to the spin equation
of motion and are therefore not verified yet. Here possible higher order corrections,
analogous to our Eq.\ (\ref{trafo}), to Eqs.\ (39) and (59) in \cite{PR08a} may be needed
to arrive at standard canonical variables for position and linear momentum.

\acknowledgments
This work is supported by the Deutsche Forschungsgemeinschaft (DFG) through
SFB/TR7 ``Gravitational Wave Astronomy''.

\end{document}